\documentclass[11pt]{article}
\usepackage[body={17.5cm, 22cm},right=2cm]{geometry}
\usepackage{cite}
\usepackage[utf8]{inputenc}
\usepackage{amsmath}
\usepackage{amssymb}
\usepackage{graphicx}
\usepackage{fullpage}
\usepackage{caption}
\usepackage{blindtext}
\usepackage{subcaption}
\usepackage{array}
\usepackage{tabularx}
\usepackage{verbatim}
\usepackage{color}
\usepackage{xcolor}
\usepackage{dsfont}
\usepackage[affil-it]{authblk}
\usepackage[bookmarks=true,pdfborder={0 0 0}]{hyperref} 
\graphicspath{{Images/}}

\newcommand{\Mpl}{M_{\rm Pl}}

\newcommand{\vev}[1]{\langle #1 \rangle}

  \def\l{\lambda}  
    \def\o{\omega}

 \newcommand{\Ocal}{{\mathcal O}}

\newcommand{\Rcal}{{\mathcal R}}

\title{Probing Bosonic Overdensities with Optomechanical Sensing}

\author{Katherine Slattery$^{a}$\thanks{Electronic address: \texttt{slattekr@mail.uc.edu}}\,\,\,}

\author{Joshua Eby$^{b,c}$\thanks{Electronic address: \texttt{joshaeby@gmail.com}}\,\,\,}

 \author{Lauren Street$^{a,d}$\thanks{Electronic address: \texttt{streetlg@mail.uc.edu}}\,\,\,}
 
 \author{Rohana Wijewardhana$^{a}$\thanks{Electronic address: \texttt{rohana.wijewardhana@gmail.com}}}
 
\affil{{ \sl $^{a}$ 
				Department of Physics, University of Cincinnati, Cincinnati, OH 45221 USA}}
\affil{{ \sl $^{b}$ 
				The Oskar Klein Centre, Department of Physics, \\ Stockholm University, 10691 Stockholm, Sweden}}
\affil{{ \sl $^{c}$ 
				Kavli Institute for the Physics and Mathematics of the Universe (WPI), \\ The University of Tokyo Institutes for Advanced Study, \\ The University of Tokyo, Kashiwa, Chiba 277-8583, Japan}}
\affil{{ \sl $^{d}$ 
				Fermi National Accelerator Laboratory, P.O. Box 500, Batavia, IL 60510, USA}}
\date{Dated: \today}

\begin{document}
\maketitle

\begin{abstract}
Previous work has shown that optomechanical force sensing can be used for efficient detection of ultralight (sub-eV) dark matter candidates. We propose to extend the reach of this method to the search for ultralight dark matter in gravitationally-bound configurations in the Milky Way. We consider three scenarios, each strongly-motivated by previous studies: boson stars traveling in the galaxy with virial velocity; a bosonic halo centered around the Sun (a `solar halo'); and a bosonic halo centered around the Earth. For each case, we consider bound states 
composed of either scalar particles with a Yukawa coupling, or vector particles coupled to baryon minus lepton number charge. 
Accounting for all experimental constraints on coupling strength, we estimate the sensitivity reach of an optomechanical sensor search.
We conclude that, although boson star encounters with Earth would be too infrequent to be detected in the relevant parameter space, current optomechanical force sensing technologies provide promising search capabilities for solar or Earth-bound halos.
\end{abstract}

\section{Introduction}

There is currently abundant astrophysical evidence for the existence of dark matter (DM), though at present no specific particle has been detected which has the required characteristics to account for it~\cite{Sofue:2000jx,ParticleDataGroup:2018ovx,Massey:2010hh,Primack:2015kpa,Planck:2018vyg}. In recent years, the search for ``ultralight" DM (ULDM) candidates, with mass below the eV scale, has intensified due to heavier candidates not being detected, despite significant efforts in a wide-ranging experimental program (see e.g.~\cite{Schumann:2019eaa} for a recent review). There are several classes of theoretically well-motivated ULDM candidates, such as axions and axion-like particles (ALPs) (see~\cite{Hui:2016ltb,DiLuzio:2020wdo,Ferreira:2020fam} for recent reviews). Specifically, both relatively heavy ULDM candidates (e.g. QCD axions~\cite{Peccei:1977hh,Weinberg:1977ma,Wilczek:1977pj,Kim:1979if,Shifman:1979if,DiVecchia:1980yfw,Zhitnitsky:1980tq,Dine:1981rt,Preskill:1982cy,Abbott:1982af,Dine:1982ah,GrillidiCortona:2015jxo}) and extremely light candidates (e.g. fuzzy dark matter~\cite{Hu:2000ke,Svrcek:2006yi,Arvanitaki:2009fg}) have been sought via both direct and indirect detection methods, which have yet to make an unambiguous detection of DM; for recent reviews of the experimental status see~\cite{Antypas:2022asj,Adams:2022pbo}.

Given the negative results for dark matter searches above, one might hope to use density fluctuations in the dark matter field to increase the interaction rate, and thereby probe a wider range of possible DM couplings to matter. 
In a widely-studied example, DM which consists primarily of such ultralight bosonic fields can condense into ``dark stars." In particular,  electrically-neutral scalar particles 
can occupy gravitationally-bound Bose-Einstein-condensed states, called ``boson stars"~\cite{Kaup:1968zz,Ruffini:1969qy,BREIT1984329,Colpi:1986ye}, which can form on astrophysical timescales in galaxies~\cite{Schive:2014dra,Levkov:2018kau,Chen:2020cef,Kirkpatrick:2020fwd,Chen:2021oot,Kirkpatrick:2021wwz}. Vector (spin-1) dark matter candidates may also form gravitationally self-bound condensates in 
the galaxy, which are called ``Proca stars"; these states have many of the same properties as scalar boson stars, as demonstrated in \cite{Adshead:2021kvl,Jain:2021pnk}, and may be produced with cosmologically-relevant abundances as well~\cite{Gorghetto:2022sue}. 

Recently, it was proposed that bosonic particles can be captured to bound states around external astrophysical bodies, such as the Sun or the Earth,  \cite{Banerjee:2019epw,Banerjee:2019xuy}; the resulting configurations are akin to ``gravitational atoms'' due to the $1/r$ Newtonian trapping potential.\footnote{These bound states from direct DM capture should be distinguished from another atom-like state, which forms by a process known as \emph{superradiance}. Superradiance produces scalar fields directly from vacuum by sapping the angular momentum of rapidly-rotating black holes~\cite{Arvanitaki:2010sy,Arvanitaki:2014wva,Arvanitaki:2016qwi,Baryakhtar:2020gao,Unal:2020jiy,Branco:2023frw} or stars~\cite{Chadha-Day:2022inf}.} Indeed, ULDM self-interactions can be sufficient to stimulate capture around the Sun and lead to overdensities (relative to the DM background) as large as $10^4$ in the vicinity of Earth~\cite{Budker:2023sex}; the overdensity can be even larger nearer to the Sun, motivating space-based direct detection experiments~\cite{Tsai:2021lly,Derevianko:2021kye,Schkolnik:2022utn}. 
Other mechanisms may lead to capture around the Earth, where other novel signals have been explored (see e.g.~\cite{Kouvaris:2021phj,Gherghetta:2023myo}). 

In the presence of such overdensities, the sensitivity of searches for DM on Earth can be significantly modified. The most important effect is 
that the DM energy density in the experiment, which is typically estimated to be $\rho_{\rm DM} = 0.4\,{\rm GeV/cm}^3$, can be much higher in the presence of a bound state in the solar system or during an encounter with a boson star. The coherent oscillations of the scalar fields in a bound state may also provide additional sensitivity compared to the background DM search; see e.g.~\cite{Budker:2013hfa,Graham:2013gfa,Banerjee:2019epw,Banerjee:2019xuy}.

The use of macroscopic force sensors to search for long-range interactions between DM and the Standard Model (SM) 
have received increased interest in recent years. Previous works have shown that existing and upcoming force-sensing technologies could offer significant 
detection reach across the ULDM parameter space~\cite{Carney:2019cio,Carney:2020xol,Manley:2020mjq,Manley:2021hao,Brady:2022qne,Baker:2023kwz}. 
In this work, we combine the two observations above, 
exploring the 
search potential for ongoing and future ULDM experiments with optomechanical sensors, focusing on scenarios with bosonic overdensities (boson stars and gravitational atoms). 

This paper is organized as follows. In Section~\ref{sec:overdensity}, we describe the overdensities (boson stars and gravitational atoms) considered in this work, focusing on their density and size. We describe the induced force on optomechanical sensing experiments in Section~\ref{sec:forces}, including a derivation from the Dirac equation, for scalar and vector DM candidates. We detail the results in Section~\ref{sec:results} and conclude in Section~\ref{sec:conclusion}. 

We work in natural units, where $\hbar = c = 1$.

\section{Bosonic Overdensities}
\label{sec:overdensity}

Consider a scalar DM field $\phi$. 
Generically the leading term of the self-interaction potential can be written as
\begin{equation}
V(\phi \phi^*) = - \frac{\lambda}{4} (\phi \phi^*)^2\,,
\end{equation}
where $\lambda$ is the self-coupling constant of the bosons; we will assume $\lambda>0$, which corresponds to an attractive self-interaction; taking a repulsive self-interaction $\lambda<0$ would leave our results essentially unchanged.\footnote{A self-interaction proportional to $\phi^3$ is also possible, but its effect is negligible in the non-relativistic limit relevant to this work.}
In the non-relativistic limit, 
the dark matter field 
can be expanded in terms of a non-relativistic wavefunction $\Psi$ as
\begin{equation}
    \phi(\vec{r},t) = \frac{1}{\sqrt{2m}}\left[\Psi(\vec{r},t) \exp(-imt) + h.c.\right]\,.
\end{equation}
As long as $\dot{\Psi} \ll m\Psi$, relativistic effects are suppressed \cite{Guth:2014hsa,Eby:2018dat}, and we may determine the macroscopic parameters of the star using the Gross-Pitaevskii equation
\begin{align} \label{eq:GP}
    i\dot{\Psi} = \left[-\frac{\nabla^2}{2m} + \Phi(\Psi) - \frac{\lambda}{8 m^2}|\Psi|^2\right]\Psi\,,
\end{align}
where the gravitational potential $\Phi = \Phi_{\rm self} + \Phi_{{\rm ext}}$ is the sum of an external gravitational potential $\Phi_{{\rm ext}}$ and the potential $\Phi_{\rm self}$ from the self-gravity of the bosons. The latter satisfies the Poisson equation
\begin{equation} \label{eq:Poisson}
    \nabla^2\Phi_{\rm self} = \frac{4 \pi m^2}{\Mpl^2}|\Psi|^2\,,
\end{equation}
where $\Mpl=1.2\times10^{19}$ GeV is the Planck mass. Eqs.~(\ref{eq:GP}-\ref{eq:Poisson}) are collectively known as the Gross-Pitaevskii+Poisson (GPP) equations.

Bosonic DM can form gravitationally-bound structures, which are BEC-like states at low temperatures~\cite{Kaup:1968zz,Ruffini:1969qy,BREIT1984329,Colpi:1986ye}. The most widely-studied case is that of spin-zero fields in self-gravitating configurations, which are generically known as \emph{(scalar) boson stars} \cite{Chavanis:2011zi,Chavanis:2011zm,Eby:2018dat,Kouvaris:2019nzd}, or \emph{axion stars} \cite{Barranco:2010ib,Eby:2014fya,Schiappacasse:2017ham}  when they are composed of pseudoscalar fields such as ALPs. Important properties of these states, including their formation~\cite{Schive:2014dra,Levkov:2018kau,Chen:2020cef,Kirkpatrick:2020fwd,Chen:2021oot,Kirkpatrick:2021wwz} and accretion rate~\cite{Eggemeier:2019jsu,Chen:2020cef,Chan:2022bkz,Dmitriev:2023ipv}, have been widely studied in recent literature. For our purposes, we will merely assume that such states can form with some total abundance, and see how one might search for them in such scenarios.

Vector (spin-1) dark matter particles may also form self-gravitating condensates, called \emph{vector boson stars} or \emph{Proca stars} \cite{Brito:2015pxa,Minamitsuji:2018kof,Amin:2019ums,Adshead:2021kvl,Jain:2021pnk,Gorghetto:2022sue}. As noted in \cite{Adshead:2021kvl}, there are different classes of self-gravitating vector solitons, including spherical, cylindrical, and planar vector boson stars. However, in the ground state, the spatial components of vector dark matter condensates have the behavior of separate, non-interacting scalar fields. Therefore, in some sense a vector boson star is equivalent to a superposition of three scalar boson stars with equal particle mass $m$. As such, we may model both scalar and vector boson star using the same formalism, e.g.
both must satisfy the GPP equations.

As noted in \cite{Eby:2018dat}, 
variational techniques are widely used as a highly-precise substitute for the full solution of the GPP equation.
For a bound state of eigenenergy $\omega$, one can write $\Psi(\vec{r},t) = \exp(i\o t)\psi(\vec{r})$, and the resulting configuration 
can be approximately determined by minimizing the energy functional derived from eq.~\eqref{eq:GP}, 
\begin{equation}
    E[\psi] = \int d^3r \bigg[ \frac{|\nabla \psi|^2}{2m}
    + \frac{m}{2} \Phi_{\rm self}|\psi|^2  
    + m \Phi_{\rm ext}|\psi|^2 - \frac{|\lambda|}{16m^2}|\psi|^4 \bigg]\,,
\end{equation} 
with regard to some ansatz for the wavefunction $\psi$. Holding the number of particles $N$ in the star fixed, the function $\psi$ must be normalized 
as
\begin{equation} \label{eq:norm}
    \int d^3r |\psi|^2 = N\,.
\end{equation}
Extrema of $E[\psi]$ correspond to (meta)stable configurations, which we analyze below. In what follows, we also assume spherical symmetry, which is appropriate for the ground-state configurations of interest here.

\subsection{Boson Stars}

For self-gravitating states with attractive self-interactions, an ansatz of the form
\begin{equation}
    \psi (r) = \sqrt{\frac{(5.4)^3 N}{7 \pi R_{99}^3}}\left(1+\frac{5.4r}{R_{99}}\right)e^{-5.4r/R_{99}} 
\end{equation}
provides an excellent approximation for the wavefunction of the star \cite{Eby:2018dat}, where $R_{99}$ is the radius containing $99\%$ of the mass of the star. In particular, this profile is cored at small $r$, and goes exponentially to $0$ as $r\to\infty$, 
matching the behavior of the exact solutions. 
Assuming $\Phi_{\rm self} \gg \Phi_{\rm ext}$ (i.e. far from any external potential), the mass of the star $M=m N$ is inversely related to the radius $R_{99}$ as\footnote{Note the difference in the definition of the size of the boson star used here compared to Ref. \cite{Banerjee:2019epw}; the latter chose a radius parameter $R$ which is related to $R_{99}$ by $R = R_{99}/5.4$.}
\begin{equation}
    R_{99} = \frac{5.4 \Mpl^2}{m^2 M}
        \left[1+\sqrt{1-\left(\frac{M}{M_c}\right)}\right]\,,
\end{equation}
where the critical (maximum) mass induced by the attractive self-interactions is
\begin{equation}
    M_c = 10.15\frac{\Mpl}{\sqrt{\l}}\,.
\end{equation}
During a boson star transit, an experiment will be subject to an increased DM flux which changes with time, but for an order-one fraction of the transit the energy density roughly equal to the central density of the boson star,
\begin{equation}
  \rho(0) \equiv m|\psi(0)|^2 
    = \frac{(5.4)^3 M}{7 \pi R_{99}^3}
    \simeq 10^4\rho_{\rm DM} 
        \left(\frac{\mu{\rm eV}}{m}\right)^2
        \left(\frac{10^8\,{\rm km}}{R_{99}}\right)^4\,.
 \end{equation}
  
A very similar calculation leads to the central density for vector boson stars as well, though the final answer will be larger by a factor of three since the vector boson stars have three polarization components. 
In what follows, we focus on the scalar case but use this simple argument to determine the density of vector boson stars as well.

\subsection{Solar Halos and Earth Halos}

 Near an astrophysical body, the bound configuration may instead satisfy $\Phi_{\rm self} \ll \Phi_{\rm ext}$, and in this case we consider a bosonic configuration bound around an external gravitational source, e.g. the Sun, another star, or a planet such as Earth. As long as $R_{99} \gg R_{\rm ext}$, the potential is well-approximated by $\Phi_{\rm ext} = G M_{\rm ext}/r$, where $M_{\rm ext}$ and $R_{\rm ext}$ are the mass and radius of the external source (respectively). The resulting configuration then resembles a gravitational Hydrogen atom, with ground-state wavefunction proportional to an exponential~\cite{Banerjee:2019epw,Banerjee:2019xuy}, and can be captured from the DM background directly via self-interactions~\cite{Budker:2023sex}. 
 
 Given the normalization of eq.~\eqref{eq:norm}, the approximate solutions are of the form~\cite{Eby:2018dat}
 \begin{equation}
     \psi(r) = \sqrt{\frac{(4.2)^3 N}{\pi R_{99}^3}}
            e^{-4.2r/R_{99}}\,,
 \end{equation} 
 where $R_{\rm 99} = 4.2a_0$ is proportional to the gravitational `Bohr radius' $a_0 \equiv (m^2 M_{\rm ext}/\Mpl^2)^{-1}$. This profile is a good approximate solution of eq.~\eqref{eq:GP} for $r\gg R_{\rm ext}$. 
 As pointed out in \cite{Banerjee:2019epw}, when $R_{99}\lesssim R_{\rm ext}$, the gravitational potential is (at leading order) proportional to $r^2$ rather than $1/r$, and the ground state instead has a Gaussian form; we will not consider such a case in this work.
 
Unlike the case of boson stars, for gravitational atoms one would not need to wait for a rare transient encounter, as the bound state would be present at all times after it forms.
The density relevant to experimental searches is
 \begin{equation}
     \rho(\Rcal) \equiv m|\psi(\Rcal)|^2\,.
 \end{equation}
 If the halo is bound to the Sun, then $\Rcal$ is identified with the distance between the Earth and the Sun; the maximum dark matter density in this case is constrained by solar system ephemerides~\cite{Anderson:1995dw,Gron:1995rn,Pitjev:2013sfa}.
 On the other hand, if the halo is bound to the Earth, then we instead identify $\Rcal$ as the distance from the center of the Earth, and the strongest constraints on the DM density arise from objects in orbit around the Earth~\cite{Adler:2008rq}.
 These constraints have been translated to the case of solar or Earth-bound halos in Ref.~\cite{Banerjee:2019epw}. Importantly, $\rho$ can be much larger than the ambient DM density $\rho_{\rm DM} \simeq 0.4\,{\rm GeV/cm}^3$, allowing the possibility of enhanced sensitivity in the presence of these bound states. 

 At present, we are not aware of any previous work on vector bosons forming gravitational-atom-like configurations bound to the Sun or Earth. However, the constraints described above should translate directly to this case, because they are derived from gravity-only observations. We will therefore treat the density of vector bound states as equivalent to the scalar case, though this may only be correct up to factors of $\Ocal(1)$. Note that the formation of vector bound states, possibly via self-interactions as in the scalar case considered in~\cite{Budker:2023sex}, is worthy of detailed study.

\section{Bosonic forces on Optomechanical Sensors}
\label{sec:forces}

Previous works have shown that optomechanical sensing could be used to search for ultralight DM \cite{Carney:2019cio,Carney:2020xol,Manley:2020mjq,Manley:2021hao,Brady:2022qne,Baker:2023kwz}. One implementation of an optomechanical force sensing device (see e.g. \cite{Carney:2019cio}) consists of an optical cavity with an interferometer, where one of the mirrors is suspended from a pendulum. This pendulum acts as the force sensor because its position can be determined by measuring the fringe shifts on the interferometer. If a DM wave interacts with the sensor, it changes the position of the sensor and consequently alters the fringe shift pattern on the interferometer. This detection scheme offers many advantages in terms of noise reduction, especially when scaled into an array of sensors \cite{Carney:2019cio}. In this paper, we consider the observable coupling which could be measured with a single sensor, though the results can be scaled up to an arbitrary number of sensors. 

We consider three benchmark DM cases below: scalars coupled to the SM through a Yukawa coupling to neutrons; axion-like particles with a parity-odd coupling to neutrons; and vector particles coupling to baryon minus lepton number (i.e. vector $B-L$ bosons). In each case, we consider the sensitivity reach in the presence of a boson star, a solar halo, and an Earth-bound halo. 

Below, we begin with a pedagogical derivation of the force on an optomechanical sensor resulting from the three bosonic DM models above.

\subsection{Force from Scalar or Pseudoscalar Couplings}
\label{sec:scalarforce}

Consider an interaction Lagrangian of the form
\begin{equation} \label{eq:Lint}
    \mathcal{L}_{\rm int} = y_1 \phi \bar{n}n
                            + y_2 \phi  \bar{n} \gamma^5 n\,,
\end{equation}
which includes a scalar Yukawa interaction $\propto y_1$
as well as a pseudoscalar interaction $\propto y_2$ 
between $\phi$ and the nucleon field $n$.
These nucleons, which make up detector apparatus (e.g. a mirror), are represented by the Dirac spinor field
\begin{equation}
	n = \begin{bmatrix}
		U \\
		V
	\end{bmatrix}.
\end{equation}
with $2$-component spinors $U$ and $V$.
This field must satisfy the Dirac equation,
\begin{equation} 
	\gamma^0 E n = [ \vec{\gamma}\cdot \vec{p} + (m + y_1 \phi)\mathds{1}+ y_2 \gamma^5 \phi] n\,,
\end{equation}
where $E$ ($\vec{p}$) is the energy (momentum) operator, $\mathds{1}$ is the identity matrix, and $\gamma^\mu$ ($\mu=0,1,2,3$) are the usual Dirac Gamma matrices.
Since $\gamma^5 = i \gamma^0 \gamma^1 \gamma^2 \gamma^3$, we may rewrite the above equation as
\begin{equation}
	E 
	\begin{bmatrix} 
		\mathds{1} & 0 
		\\0 & -\mathds{1}
	\end{bmatrix}
	\begin{bmatrix}
		U \\ 
		V
	\end{bmatrix} = 
	\begin{bmatrix} 
		0 & \vec{\sigma}\cdot \vec{p} \\ -\vec{\sigma}\cdot \vec{p} & 0
	\end{bmatrix}
	\begin{bmatrix}
		U\\
		V
	\end{bmatrix} +
\begin{bmatrix}
	y_1 \phi U \\
	y_1 \phi V
\end{bmatrix} +
	m \begin{bmatrix}
		U \\
		V
	\end{bmatrix} + y_2 \phi \begin{bmatrix}
		V \\
		U
	\end{bmatrix}\,,
\end{equation} 
where $\vec{\sigma}$ is the Pauli vector.
Therefore the following pair of equations must be simultaneously satisfied:
\begin{align}
	E U &= (\vec{\sigma}\cdot \vec{p})V + m U + y_1 \phi U+  y_2\phi V \,,\\
    -E V &= -(\vec{\sigma}\cdot \vec{p})U + m V +y_1\phi V + y_2 \phi U\,.
\end{align}

Solving the above equations for $U$ and $V$ yields
\begin{align} \label{eq:vNR}
	V &= \frac{1}{2 m + y_1 \phi} (\vec{\sigma}\cdot\vec{p} - y_2 \phi)U\,, \\
    (E-m -y_1\phi) U &= \frac{1}{2m + y_1 \phi} [ (\vec{\sigma}\cdot\vec{p})^2 + (y_2\phi)(\vec{\sigma}\cdot\vec{p})- (\vec{\sigma}\cdot \vec{p})(y_2 \phi)- y_2^2\phi^2]U \,. \label{eq:uNR}
\end{align}
We will consider the effect of the two couplings $y_1$ and $y_2$ separately below.

\begin{itemize}
\item \emph{Case 1 (scalar, $y_2 = 0$):}
Taking $K \equiv E -m$, the preceding equation becomes
\begin{equation}
	K U = y_1 \phi\,U + \frac{(\vec{\sigma}\cdot\vec{p})^2}{2m +y_1 \phi} U\,.
\end{equation}
Expanding the denominator to first order in powers of $y_1\phi/m$, we have\footnote{Given the smallness of the couplings we consider, this approximation works extremely well.}
\begin{equation}
	K U \simeq y_1 \phi\,U 
        + \frac{\left(\vec{\sigma} \cdot \vec{p}\right)^2}{2m}\left(1-\frac{y_1 \phi}{2m}\right)
        \,U 
        \simeq  [y_1 \phi + \frac{p^2}{2m}] \,U \,,
\end{equation}
where we used $(\vec{\sigma}\cdot \vec{p})^2 = p^2$.
Identifying $p^2/2m$ as the kinetic energy of a free particle, the force on a particle in the mirror during an interaction with DM is therefore 
\begin{equation}
	F_1 \equiv -\nabla\left(K-\frac{p^2}{2m}\right) = -y_1 \nabla \phi\,.
\end{equation}
If the number of SM fermions in the mirror is $N_g$, the total force on the mirror will be $F=N_g F_1$ with
\begin{equation} \label{eq:Fscalar}
    F = y_1 N_g v \sqrt {2\rho} \sin(\omega_{\phi} t),
\end{equation}
where we used 
$\phi=\frac{\sqrt{2\rho}}{m} \cos (\vec{p}\cdot\vec{x}-\omega_\phi t)$. Note that we dropped the $x$-dependence because in what follows, the wavelength of $\phi$ will always be much larger than the size of the experiment.

\item \emph{Case 2 (pseudoscalar, $y_1 = 0$):} 
In this case, eqs.~(\ref{eq:vNR}-\ref{eq:uNR}) simplify to
\begin{align}
	V &= \frac{1}{2 m} (\vec{\sigma}\cdot\vec{p} - y_2 \phi)U\,, \\
    (E-m) U &= \frac{1}{2m} [ (\vec{\sigma}\cdot\vec{p})^2 + (y_2\phi)(\vec{\sigma}\cdot\vec{p})- (\vec{\sigma}\cdot \vec{p})(y_2 \phi)- y_2^2\phi^2]U\,.
\end{align}
Since the Pauli spin matrices commute with the momentum operator and since $[\phi, p_i] = i \partial_i \phi$, we may rewrite the preceding equation as
\begin{equation}
	(E-m)U = \frac{1}{2m}[p^2+ y_2 i (\vec{\sigma}\cdot\vec{\nabla}\phi)- y_2^2 \phi^2]U\,.
\end{equation}
As before, we define $K\equiv E-m$
and find the associated (spin-dependent) force $-\nabla\left(K-p^2/2m\right)$, evaluated to 
\begin{equation}
	\hat{F} = - \nabla\left[-\frac{y_2^2}{2m} \phi^2 + i\frac{y_2}{2m}(\vec{\sigma}\cdot \nabla\phi)\right]\,.
\end{equation}
Dropping the (small) $y_2^2$ term, we obtain 
\begin{equation}
\hat{F} \simeq 	-\frac{iy_2}{2m} \vec{\nabla}\left[(\vec{\sigma}\cdot \vec{\nabla}\phi)\right]\,,
\end{equation}
Note that unlike the case in eq.~\eqref{eq:Fscalar}, $\hat{F}$ here (denoted here with a `hat') should be thought of as an operator acting on the nucleon spin. 
\end{itemize}

A well-motivated example which is classified under Case 2 is a derivative coupling between an axion field $\phi$ and 
a nucleon field $\xi$ of the form 
$(g/f) \partial_\mu \phi (\bar{\xi}\gamma^\mu \gamma^5 \xi)$, where $g$ is a dimensionless coupling and $f$ is the axion decay constant.
The corresponding action is given by
\begin{equation}
    S = \int d^4x  \frac{g}{f} \partial_\mu \phi (\bar{\xi}\gamma^\mu \gamma^5 \xi) 
            = -\frac{g}{f}\int d^4x  \phi \partial_\mu (\bar{\xi} \gamma^\mu \gamma^5 \xi)\,,
\end{equation}
where the right-hand side of the above equation was obtained by integrating by parts and enforcing that $\phi$ must vanish at infinity. 
The integrand of $S$ then simplifies to
\begin{equation}
    \phi\partial_\mu (\bar{\xi}\gamma^\mu \gamma^5 \xi) 
            = \phi\left(\bar{\xi}\gamma^\mu \gamma^5 \partial_\mu \xi + \partial_\mu \bar{\xi} \gamma^\mu \gamma^5 \xi\right)
            = -\phi\bar{\xi} \gamma^5 \gamma^\mu \partial_\mu \xi
            = -2 \phi\bar{\xi}\gamma^5 \left(\frac{m_\xi}{i} \xi\right) 
            = 2 im_\xi \phi \bar{\xi}\gamma^5\xi\,.
\end{equation}
Combining the two previous equations yields
\begin{equation}
	S  = -\frac{2 m_\xi g}{f}\int d^4x 
                \phi i \bar{\xi}\gamma^5\xi\,,
\end{equation}
which corresponds to the second term of eq.~\eqref{eq:Lint} with $y_2 = 2m_\xi g/f$.
Since the prefactor is proportional to $m_\xi/f \ll 1$, this force is strongly suppressed; we will therefore focus on the other two cases (scalar Yukawa and vector $B-L$ forces) in what follows.

\subsection{Force from Coupling to Vector $B-L$ Bosons}

We consider a spin-1 particle which couples to baryon minus lepton number, $B-L$. Writing the equation of motion of the nucleon field $n$ in the non-relativistic limit, we may determine the (dark) electric force associated with this coupling, which is analogous to the Coulomb force on an electron coupling to photons (see~\cite{Ryder:1985wq} for details). Vector particles can also couple to ordinary electric charge, but optomechanical sensors in our chosen detection scheme (see above) are charge-neutral, and therefore would not be sensitive to 
a direct charge coupling. 

Following the discussion in \cite{Carney:2019cio}, a good target for a spin-$1$ field $A'_\mu$ is an interaction of the form
\begin{equation}
	\mathcal{L} \supset i g_{B-L} A'_\mu \bar{n} \gamma^\mu n\,,
\end{equation}
where $g_{B-L}$ is the coupling strength.
After a very similar derivation that we saw in Section~\ref{sec:scalarforce}, one obtains the force on the sensor as
\begin{equation} \label{eq:Fvector}
	F = N_{B-L} g_{B-L} \sqrt{2\rho} \sin(\omega_\phi t)
\end{equation}
where $N_{B-L}$ is the $B-L$ charge of the sensor; note that for charge-neutral materials, this reduces to the number of neutrons in the sensor. Relative to the scalar Yukawa case in eq.~\eqref{eq:Fscalar}, the main difference is that the force from a vector is not suppressed by $v\simeq 10^{-3}$. This is because the vector force is proportional to the (dark) electric field, so the result does not depend on the derivative of the vector field.

\section{Results}
\label{sec:results}

\subsection{Previous Constraints}
\label{ssec:BBN}

There are strong constraints on ultralight scalar couplings to the SM 
arising from searches for violations of the Equivalence Principle (EP)~\cite{Hees:2018fpg,KONOPLIV2011401,Fischbach:1996eq,Adelberger:2003zx,Berge:2017ovy}. Other direct searches, including those using atomic clocks and interferometry, are summarized in Ref.~\cite{Antypas:2022asj} (see references therein); for nucleon couplings, these searches are most sensitive at low masses $m\lesssim 10^{-18}\,{\rm eV}$ and are therefore complimentary to those considered here.

One can also constrain the presence of DM coupling to neutrons through its effect on the production of primordial nuclei predicted by Big Bang Nucleosynthesis (BBN) \cite{Blum:2014vsa}. 
In particular, the observed helium abundance places a tight constraint on the Yukawa coupling. At the BBN epoch, the ratio of neutron to proton abundances is given by 
 \begin{equation} \label{eq:NnNp}
     \frac{N_n}{N_p} = e^{-Q/T_F}\,,
 \end{equation} 
 where $Q\equiv m_n - m_p \simeq 1.293\,{\rm MeV}$ is the neutron-proton mass difference and $T_F=0.8\,{\rm MeV}$ is the average temperature of the universe at freeze-out.  The mass fraction of Helium is related to $N_n/N_p$ as 
 \begin{equation}
     Y_p \approx \frac{2(N_n/N_p)}{1+(N_n/N_p)}\,.
 \end{equation}
 
 The presence of a Yukawa coupling $y_1$ to a nucleon shifts its mass proportionally to $y_1$; see eq.~\eqref{eq:Lint}. If the coupling to protons and neutrons is asymmetric, it therefore contributes to the mass difference $Q$. In the extreme case, where $\phi$ couples only to neutrons, the resulting shift is $Q\to Q+\delta Q$ with $\delta Q=y_1\vev{\phi}$, with $\vev{\phi}$ the vacuum expectation value (vev) of $\phi$. 
 Through eq.~\eqref{eq:NnNp}, this leads to a modification of the neutron-proton ratio, which at leading order is given by
 \begin{equation}
  \frac{\delta(N_n/N_p)}{N_n/N_p} \simeq 1 - \frac{\delta Q}{T_F}\,.
 \end{equation}
 The resulting fractional change in $Y_p$ is
\begin{equation}
     \frac{\delta Y_p}{Y_p} \simeq \left(\frac{2 \delta(N_n/N_p)}{[1+(N_n/N_p)]^2}\right)\left(\frac{1+N_n/N_p}{2N_n/N_p}\right) 
        = \frac{\delta(N_n/n_p)}{N_n/N_p(1+N_n/N_p)}
    \simeq \frac{1-\delta Q/T_F}{1+e^{-Q/T_F}}\,.
 \end{equation}

Since the observed helium abundance is known within $10\%$ accuracy \cite{ParticleDataGroup:2020ssz}, we take  $\delta Y_p/Y_p< 0.1$ as a conservative constraint. 
Taking $\vev\phi \simeq \sqrt{2\rho_{\rm BBN}}/m$, with $\rho_{\rm BBN}$ the dark matter energy density during BBN, this implies a constraint $y_1 \lesssim 4\cdot10^{-4}\left(m/{\rm eV}\right)$. 
Explicitly, we use  
\begin{equation}
    \rho_{\rm BBN} = \rho_{\rm MRE}\left(\frac{5\cdot10^9}{3400}\right)^4 = \rho_{\rm today}\left(\frac{3400}{1}\right)^3\left(\frac{5\cdot10^9}{3400}\right)^4\,.
\end{equation}
with $\rho_{\rm today} = 10^{-6}\,{\rm GeV/cm}^3$~\cite{Pospelov:2010hj}.

\subsection{Transient Signals from Boson Star Encounters}
\label{ssec:resultsBS}

We now turn to the signals in optomechanical sensors from boson stars. 

As noted in \cite{Kouvaris:2019nzd}, scalar DM candidates with attractive self-interactions give rise to objects with lower compactness ratios $C = M/R_{99}$ than those with repulsive (or no) self-interactions. 
Such stars would have a greater probability of passing through the Earth, 
giving an increased likelihood that an optomechanical sensing search will ``see'' a boson star encounter. On the other hand, lower compactness also implies a weaker signal in the event of an encounter, since the strength of the signal is proportional to $\phi\propto \psi \sim \sqrt{2\rho/m}$, where $\rho$ is the mass density of the star (see Section \ref{sec:forces}). This tradeoff implies a potentially limited range of viable boson star parameters for such a search; see e.g. \cite{Banerjee:2019epw}.

We estimate the encounter rate by calculating the cross-section between the detector and a boson star as
 \begin{equation}
    \sigma \approx \pi R_{99}^2\,,
\end{equation}
where we have assumed the boson star radius $R_{99}$ is much larger than the detector (this is always the case in the parameter space we explore here).
The mean free path is $L = (n \sigma)^{-1}$ where $n$ is the number density of boson stars. 
Under the assumption that dark matter 
consists of a fraction $f$ of boson stars of mass $M$, their number density is given by 
\begin{equation}
   n = \frac{f\rho_{\rm DM}}{M} 
    \approx 10^{-16} f\,R_E^{-3}\left(\frac{10^{15}\,{\rm kg}}{M}\right)\,,
\end{equation}
where $R_E=6371\,{\rm km}$ is the radius of the Earth. 
The resulting frequency of encounters between Earth and boson stars is
\begin{equation}
    \Gamma = \frac{v}{L} \approx 0.2 f\, {\rm yr}^{-1}
        \left(\frac{\mu {\rm eV}}{m}\right)
        \left(\frac{R_{99}}{10^8\,{\rm km}}\right)^3\,,
\end{equation}
where we assumed the virial velocity $v \simeq 10^{-3}$ for the boson stars.

Ref. \cite{Kouvaris:2019nzd} determined along these lines that for $m\simeq 10^{-8}-10^{-4}$ eV and $\lambda \simeq 10^{-46}-10^{-42}$, the encounter rate can be greater than $1$/year, allowing (in principle) for direct searches for transient signals. 
However, this range of masses correspond to signals in the sensor with frequencies on the order of GHz or larger. In general, optomechanical sensors of the kind considered here have limited sensitivity to such high frequencies, in part due to measurement-added (e.g. back-action) noise; see~\cite{Carney:2019cio} for details. We therefore conclude that, for optomechanical sensors, signals from transient boson stars are likely to be either rare or too weak to detect.

\subsection{Detection Reach in the Presence of Solar and Earth Halos}
\label{ssec:resultsHalos}

Finally, we turn to the case of gravitational atoms bound to the Sun or Earth. If such states form, their density can be much larger than the DM background and they would be expected to remain approximately static on astrophysical timescales (see e.g.~\cite{Budker:2023sex} for details). Therefore it is a prime target for direct searches, e.g. in optomechanical sensors. 

For a scalar particle, the force on an optomechanical sensor is proportional to the relative velocity between the dark matter field (taken to be $v\simeq 10^{-3}$), as well as the square root of the dark matter density $\rho$, as in eq.~\eqref{eq:Fscalar}. Therefore to estimate the sensitivity, we use the scaling relationship
\begin{equation} \label{eq:y1sens}
    \left(y_1\right)_{\rm limit} \propto 
        \frac{1}{F} \propto \frac{1}{v \sqrt{\rho}}\,.
\end{equation}
The sensitivity for a vector particle is similar, but without the factor of $1/v$; see eq.~\eqref{eq:Fvector}. In this case one has
\begin{equation}
    \left(g_{B-L}\right)_{\rm limit} \propto 
        \frac{1}{F}\propto \frac{1}{\sqrt{\rho}}\,.
\end{equation}

Importantly, the velocity dispersion in a bound halo is much smaller than $v_{\rm dm}=10^{-3}$, which is typical of the DM background; see discussion in~\cite{Banerjee:2019xuy}. There are two components to the velocity dispersion in a bound halo: a \emph{radial} component, $v_{\rm rad}\equiv \nabla_r\phi/(m\,\phi) \simeq (m  R_\star)^{-1}$ arising from the gradient of the wavefunction; and a \emph{tangential} component, $v_{\rm tan}\equiv v_{\rm rel}$ arising from the relative velocity $v_{\rm rel}$ of the detector through the halo. We will assume a static halo for simplicity; then, for a solar halo, $v_{\rm rel}$ is of order $v_\odot\simeq 10^{-4}$ (the speed of the Earth around the Sun), and for an Earth halo it is of order $v_\oplus\simeq 10^{-6}$ (the rotation speed of Earth at the Equator). In our sensitivity estimates for the scalar field, we will use $v={\rm max}\left[(m R_\star)^{-1},v_{\rm rel}\right]$, which captures the size of the effect but ignores the difference in the direction of the force (which is in the direction of $\vec{v}$, as shown by eq.~\eqref{eq:Fscalar}). 

For simplicity, in deriving the sensitivity we assume that the sensor has the characteristics described in Section~\ref{sec:scalarforce}, which follows the discussion in~\cite{Matsumoto:2018via,Carney:2019cio}. 
Using the maximal dark matter density allowed by astronomical data, as discussed in~\cite{Banerjee:2019epw},  
we rescale the results of~\cite{Carney:2019cio} by the factor $\sqrt{\rho/\rho_{\rm DM}}$ to estimate the sensitivity reach in the presence of a bound state. For scalars, we also include the important factor of $v$ in eq.~\eqref{eq:y1sens}. 

\begin{figure}[!t]
	\centering
\includegraphics[scale=0.86]{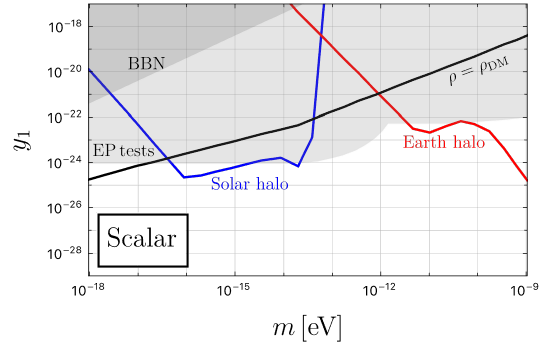} \quad
\includegraphics[scale=0.86]{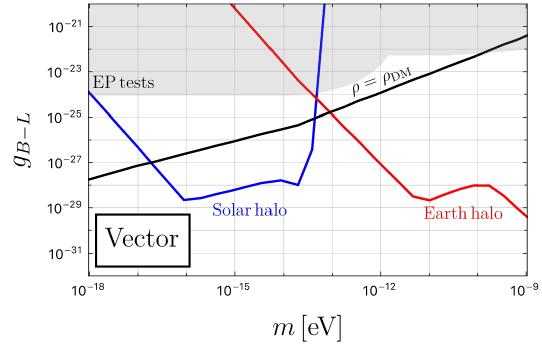}
	\caption{Sensitivity estimate for optomechanical sensor searches for ultralight scalars (left) and vectors (right). The black lines represent the estimate of Carney et al.~\cite{Carney:2019cio} using the local DM density $\rho_{\rm DM}=0.4\,{\rm GeV/cm}^3$; the blue and red regions are our estimations for a similar experimental search in the presence of a bound state around the Sun or Earth (respectively). The light grey regions represent existing limits from tests of the Equivalence Principle (EP)~\cite{Hees:2018fpg,KONOPLIV2011401,Fischbach:1996eq,Adelberger:2003zx,Berge:2017ovy}, and the dark grey in the left panel represents the constraint from Big Bang Nucleosynthesis (BBN) (see Section~\ref{ssec:BBN}).
 }
	\label{fig:OptoMech}
\end{figure}

The resulting sensitivity estimates are shown in Figure~\ref{fig:OptoMech}, for scalars (left) and vectors (right). We observe the expected increase in sensitivity at small couplings in both cases, though importantly for the scalar case, the enhancement is sufficient to probe novel parameter space beyond what is tested by EP tests (light gray regions~\cite{Hees:2018fpg,KONOPLIV2011401,Fischbach:1996eq,Adelberger:2003zx,Berge:2017ovy}). We also show for comparison the BBN constraint on scalar Yukawa interactions in the top-left corner of the left panel (dark grey).

Note that the sensitivity is estimated using the \emph{maximum} DM density allowed by current constraints (see~\cite{Banerjee:2019epw}). Recent work~\cite{Budker:2023sex} shows that such large densities (up to $\rho/\rho_{\rm DM}\sim 10^4$) are achievable for solar halos with scalar masses near $10^{-14}\,{\rm eV}$, through capture by self-scattering of scalar particles; other mechanisms would be required to capture a large density of scalars for other parameters and/or around the Earth.
 Note also that this sensitivity estimates neglect self-interactions of the ULDM, which affect the stability of the bound state and could ultimately lead to smaller bound densities. As discussed above, in this work we remain agnostic as to the source of the overdensity around the Sun or Earth, and focus on the phenomenology of the bound state and its impact on future optomechanical sensing searches. 

This experimental search described here is constrained by the noise profile, as described in~\cite{Carney:2019cio}; for completeness we briefly summarize below.
The noise in this experiment comes from thermal effects and measurement-added sources. Measurement-added noise is a quantum mechanical consequence of the measurement itself, arising from both ``shot noise" and ``back-action" noise. Shot noise refers to refers to random fluctuations in laser phase, leading to noise in the interferometer readout; while back-action noise refers arises from random fluctuations in laser amplitude which causes randoms forces to be exerted on the sensor. In addition, the specific sensor protocol used also impacts detection reach. Since the magnitude of both shot noise and back-action noise are dependent on laser power, the target dark matter frequency, and the mechanical damping of the sensor, scanning over a range of laser powers and mechanical frequencies allows one to achieve the fundamental limit on detection reach. For each case, we assume sensor characteristics as described in \cite{Matsumoto:2018via} and a sensor protocol where both the laser power and the mechanical frequency are scanned over a wide range to achieve sensitivities at the standard quantum limit. 

The dark matter oscillates with a frequency $\omega_{\phi}\simeq m$, which is coherent on a timescale $T_{\rm coh} \sim 2\pi(m v^2)^{-1}$; in the presence of a halo, this coherence time should be much longer (see discussion in~\cite{Banerjee:2019xuy}). 
This coherence time is critical to the sensitivity of the experiment: If the experiment is operating for longer that the coherence time, then the direction of the force on the sensor may change direction during the measurement. Since the timescale of this experiment is constrained by technical factors such as laser stability, it will be operated for up to several hours. For high frequency dark matter candidates, one run of this experiment will obtain a coherent signal; for lower frequency candidates, the data will have to be taken in bins which are summed in quadrature. In either case, we expect the experimental timescale to be much smaller than $T_{\rm coh}$, and therefore we ignore this effect in the estimations above.

\section{Conclusion}
\label{sec:conclusion}

We have studied the phenomenology of bound states of bosonic DM candidates in optomechanical sensing experiments. The most promising coupling types considered were a scalar Yukawa interaction and a vector $B-L$ coupling to nucleons. 
Pseudoscalar couplings do, in principle, also give rise to novel forces on these systems as well, but the signal is expected to be very suppressed.

We focused on three classes of bound states: boson stars; solar-bound halos; and Earth-bound halos. 
Boson stars could pass through an experiment with a large density, boosting its sensitivity for a finite time. However, we find that such transits are very rare except for relatively large $m\gtrsim 10^{-8}\,{\rm eV}$, where optomechanical sensors have diminished sensitivity. 

However, we find that bound states of bosonic particles around the Earth or Sun are a good target for future searches. The main advantages in the presence of a bound halo are that the density in such bound state can be orders of magnitude higher than the ambient DM density, and that the bosons oscillate coherently over very long timescales, leading to enhanced sensitivity for searches over long integration times. Importantly, in the presence of such a bound state, optomechanical sensing searches can probe novel parameter space for both vector $B-L$ interactions as well as scalar Yukawa interactions, in both cases reaching below existing limits from EP tests.

\section*{Acknowledgements}

We thank Daniel Carney for helpful discussions. K.S. thanks the University of Cincinnati Physics Department and the WISE program for funding. The work of JE was supported by the World Premier International Research Center Initiative (WPI), MEXT, Japan and by the JSPS KAKENHI Grant Numbers 21H05451 and 21K20366, as well as by the Swedish Research Council (VR) under grants 2018-03641 and 2019-02337.  
L.S. was supported by the U.S. Department of Energy (DOE), Office of Science, Office of Workforce Development for Teachers and Scientists, Office of Science Graduate Student Research (SCGSR) program. The SCGSR program is administered by the Oak Ridge Institute for Science and Education (ORISE) for the DOE. ORISE is managed by ORAU under contract number DE-SC0014664. All opinions expressed in this paper are the
authors’ and do not necessarily reflect the policies and views of DOE, ORAU, or ORISE.
Research of L.C.R.W. is partially supported by the US. Department of Energy grant DE-SC1019775.
This article is based upon work from COST Action COSMIC WISPers CA21106, supported by COST (European Cooperation in Science and Technology).

\bibliographystyle{unsrt}
\bibliography{references}

\end{document}